\documentstyle[aps,twocolumn,epsf,prb,floats]{revtex}
\def\geqap{\,\raise 2pt \hbox{$>\kern-11pt \lower 5pt \hbox{$\sim$}$}\,}
\def\leqap{\,\raise 2pt \hbox{$<\kern-10pt \lower 5pt \hbox{$\sim$}$}\,}
\begin{document}
\draft
\twocolumn[\hsize\textwidth\columnwidth\hsize\csname @twocolumnfalse\endcsname
\title{Phase Transition in Perovskite Manganites with Orbital Degree of Freedom}
\author{S.~Okamoto, S.~Ishihara, and S.~Maekawa}
\address{Institute for Materials Research, Tohoku University, Sendai 980-8577, Japan}
\date{\today}
\maketitle
\begin{abstract} 
Roles of orbital degree of freedom of Mn ions in phase transition 
as a function of temperature and hole concentration 
in perovskite manganites are studied.  
It is shown that  
the orbital order-disorder transition is of the first 
order in the wide region of hole concentration and the N$\rm \acute{e}$el temperature for the anisotropic spin 
ordering, such as the layer-type antiferromagnetic one, 
is lower than the orbital ordering temperature 
due to the anisotropy in the orbital space. 
The calculated results of the temperature dependence of 
the spin and orbital order parameters 
explain a variety of the experiments observed in manganites.
\end{abstract}
]
\narrowtext
\noindent
\section{Introduction}
Perovskite manganites $A_{1-x}B_x$MnO$_3$ ($A$: La, Nd, Pr, $B$: Sr, Ca)  
and their related compounds have been recently 
studied extensively from both experimental and theoretical sides. 
Anomalous electric, magnetic and structural properties 
accompanied with the phase transition such as colossal 
magnetoresistance (CMR) attract much attention. \cite{chahara,helmolt,tokura,jin}
Gigantic decrease of the electrical resistivity 
is brought about at the vicinity of the transition from 
spin, charge and orbital ordered phase to ferromagnetic metallic 
one with slightly changing temperature and/or applying external fields. \cite{tomioka,schiffer} 
It is now accepted that such dramatic phenomena 
are not understood within the simple double exchange scenario. \cite{millis}
\par
The orbital degree of freedom in Mn ions is one of the convincing candidates 
to bring about not only a rich variety of phenomena but also dramatic ones. 
Due to the strong Hund coupling and the crystalline field, 
two $e_g$ orbitals of a Mn ion, i.e. the $d_{3z^2-r^2}$ and $d_{x^2-y^2}$ 
orbitals, are degenerate and one of them is occupied by an electron in a Mn$^{3+}$ ion. 
It is well known that the $(d_{3x^2-r^2}/d_{3y^2-r^2})$-type orbital 
ordered state, where 
the two orbitals are alternately aligned, is realized in the 
undoped manganites LaMnO$_3$. \cite{goodenough,kanamori,matsumoto,rodriguez,murakami2}
After the discovery of CMR,  
much study of the orbital states in doped manganites has been done.
For example, the layer (A)-type antiferromagnetic (AF) metal 
accompanied with the uniform alignment of the $d_{x^2-y^2}$ orbital 
is observed in Pr$_{0.5}$Sr$_{0.5}$MnO$_3$, 
Nd$_{0.45}$Sr$_{0.55}$MnO$_3$ \cite{kawano,kuwahara,tomioka2} 
and La$_{1-x}$Sr$_x$MnO$_3$ with $x \sim 0.55$. \cite{akimoto}
Nevertheless, roles of the orbital in phase transition 
are still far from our understanding. 
Since CMR appears near the orbital order-disorder transition, 
it is indispensable to study nature of the phase transition 
in doped manganites where the orbital degree of freedom is taken into account. 
\par
In this paper, we study the phase transition in perovskite manganites 
based on the model where the orbital degree of freedom 
and electron correlation are included.
By adopting the mean field theory, 
the phase transitions are investigated as a functions of temperature ($T$) and carrier concentration ($x$). 
Since there is a strong anisotropy in the orbital space unlike the spin one,  
the orbital order-disorder transition is of the first order in the wide range of $x$, 
and the N${\rm \acute e}$el temperature $T_N$ for the 
anisotropic spin ordering, such as the A-type AF one, is lower than 
the orbital ordering temperature $T_{OO}$. 
The calculated temperature dependence 
of the spin and orbital order parameters explains a variety of experiments 
in several manganites. 
\par
In Sect.~II, the model Hamiltonian is derived and the mean field theory at finite $T$
is introduced. 
In Sect.~III, the numerical results of the spin and orbital phase diagram 
are presented. We focus on the phase transitions in (i) lightly doped region where 
the ferromagnetic Curie temperature, $T_{C}$, and $T_{OO}$ are close with each other and (ii)
highly doped one where the A-type AF state accompanied with the 
$d_{x^2-y^2}$ orbital appears. 
In Sect.~IV, phase transition is studied analytically by expanding 
the free energy with respect to the spin and orbital order parameters. 
Section V is devoted to the summary and discussion. 
\section{Model and formulation} 
Let us set up the model Hamiltonian describing the electronic structure 
in manganites.  
We consider the tight-binding 
Hamiltonian in the cubic lattice consisting of Mn ions. 
At each site, two $e_g$ orbitals are introduced  
and $t_{2g}$ electrons are treated as a localized spin ($\vec S_{t_{2g}}$) with $S=3/2$.
We introduce three kinds of Coulomb interaction between $e_g$ 
electrons at the same site, i.e. 
the intra-orbital Coulomb interaction $(U)$, the inter-orbital one $(U')$ and 
the exchange interaction $(I)$. 
The hopping integral 
between site $i$ with orbital $\gamma$ 
and its nearest neighboring site $j$ with $\gamma'$ 
is denoted by $t_{ij}^{\gamma \gamma'}$.
The Hund coupling ($J_H$) between $e_g$ and $t_{2g}$ spins 
and the antiferromagnetic superexchange (SE) interaction ($J_{AF}$) between 
nearest neighboring $t_{2g}$ spins are also introduced. 
Among these parameters, 
the intra-site Coulomb interactions are the largest.  
Thus, by excluding the doubly occupied state in the $e_g$ orbitals, 
we derive the effective Hamiltonian describing the low energy 
electronic states:\cite{ishihara1} 
\begin{eqnarray}
{\cal H}= {\cal H}_t + {\cal H}_J + {\cal H}_H + {\cal H}_{AF}.
\label{eq:heff}
\end{eqnarray}
The first and second terms correspond to the so-called 
$t$- and $J$-terms in the $tJ$ -model and are given by  
\begin{eqnarray}
{\cal H}_{t}
= \sum_{\langle ij \rangle \gamma \gamma' \sigma} 
t_{i j}^{\gamma \gamma'} \widetilde{d}_{i \gamma \sigma}^{\dagger}
\widetilde{d}_{j \gamma' \sigma} + H.c. \, ,
\label{eq:ht}
\end{eqnarray}
and 
\begin{eqnarray}
\lefteqn{{\cal H}_{J}} \nonumber \\
&=&-2J_1\sum_{\langle ij \rangle } 
 \biggl ( {3 \over 4} n_i n_j + \vec S_i \cdot \vec S_j   \biggr )
 \biggl ( {1 \over 4}  - \tau_i^l \tau_j^l \biggr ) \nonumber \\
&&-2J_2\sum_{\langle ij \rangle } 
 \biggl ( {1 \over 4} n_i n_j  - \vec S_i \cdot \vec S_j   \biggr )
 \biggl ( {3 \over 4}   + \tau_i^l \tau_j^l +\tau_i^l+\tau_j^l \biggr ) ,
\label{eq:hj}
\end{eqnarray}
respectively. 
Here, $\tau_i^l = \cos (\frac{2\pi}{3}m_l) T_{iz} - \sin (\frac{2\pi}{3}m_l) T_{ix}$ 
and $(m_x, m_y, m_z) = (1, -1, 0)$.  
$l$ denotes a direction of the bond connecting site $i$ and site $j$.  
$\widetilde{d}_{i \gamma \sigma}$ is the annihilation operator of the $e_g$ electron 
at site $i$ with spin $\sigma$ and orbital $\gamma$ with excluding double occupancy 
of electron and 
$n_i$ is the number operator defined as 
$n_i =\sum_{\gamma \sigma} \widetilde d_{i \gamma \sigma}^\dagger \widetilde d_{i \gamma \sigma}$.
The explicit form of 
$t_{i j}^{\gamma \gamma'}$ is determined by the Slater-Koster formulas. \cite{slater}
$\vec{S}_i$ is the spin operator of the $e_g$ electron with $S=1/2$ and 
$\vec{T}_i$ is the pseudo-spin one for the orbital degree of freedom defined as 
$\vec{T}_i = (1/2) \sum_{\gamma \gamma' \sigma} 
\tilde{d}_{i \gamma \sigma}^{\dagger}(\vec{\sigma})_{\gamma \gamma'}\tilde{d}_{i \gamma' \sigma}$ 
where 
$T_{iz} = +(-)1/2$ corresponds to the state where the $d_{3z^2-r^2}$ ($d_{x^2-y^2}$) orbital 
is occupied by an electron.  
$J_1 = t_0^2/(U' - I)$ and $J_2 = t_0^2/(U' + I + 2J_H)$ 
where $t_0$ is the hopping integral between nearest neighboring 
$d_{3z^2-r^2}$ orbitals in the $z$ direction and $U = U' + I$ is assumed.  
The third and fourth terms in Eq.~(\ref{eq:heff}) represent 
the Hund coupling and the antiferromagnetic SE interaction, respectively, 
and are given by 
\begin{eqnarray}
{\cal H}_{H} + {\cal H}_{AF}
= &-&J_H \sum_i \vec{S}_{i} \cdot \vec{S}_{t_{2g} i} \nonumber \\
&+&J_{AF} \sum_{\langle i j \rangle} \vec{S}_{t_{2g} i} \cdot \vec{S}_{t_{2g} j} . 
\label{eq:hhhaf}
\end{eqnarray}
The detailed derivation of the Hamiltonian is presented in Ref.~\onlinecite{ishihara1}. 
Characteristics of this Hamiltonian are summarized as follows;  
1) the two kinds of magnetic interactions between spins of $e_g$ electrons, i.e. 
the SE and double exchange 
interactions are described by ${\cal H}_J$ and ${\cal H}_t$, 
respectively, \cite{endoh}
2) there is a strong anisotropy in the pseudo-spin space 
unlike the spin space, and 
3) the first term in Eq.~(\ref{eq:hj}) is the dominant term in $ {\cal H}_J $ and 
stabilizes the ferromagnetic state associated with the antiferro-type 
orbital ordered one where different types of orbital are alternately 
aligned. \cite{roth,kugel,cyrot,inagaki}
\par
Being based on the Hamiltonian, 
the spin and orbital states are studied at finite $T$ and $x$. 
The mean field theory proposed by de Gennes \cite{degennes}
is applied to the present system with orbital degeneracy and 
electron correlation. \cite{okamoto}
In this theory, the spin and 
pseudo-spin operators are treated as classical vectors: 
\begin{eqnarray}
(S_{ix}, S_{iy}, S_{iz}) = 
{1 \over 2} (\sin \theta_i^s \cos \phi_i^s, \sin \theta_i^s \sin \phi_i^s, 
\cos \theta_i^s)\label{eq:sdef}, 
\end{eqnarray}
and
\begin{eqnarray}
(T_{ix}, T_{iy}, T_{iz}) = {1 \over 2} (\sin \theta_i^t, 0, \cos \theta_i^t)\label{eq:tdef},
\end{eqnarray}
where $\theta_i^{s(t)}$ is the polar angle in the spin (orbital) space 
and $\phi_i^s$ is the azimuthal one in the spin space. 
A motion of the pseudo-spin is assumed to be confined in the $xz$ plane and  
$\theta_i^t$ describes the orbital state at site $i$ as follows 
\begin{eqnarray}
| \theta_i^t \rangle = 
\cos {\theta_i^t  \over 2} | 3z^2-r^2 \rangle + \sin {\theta_i^t  \over 2} | x^2-y^2 \rangle .
\label{eq:thetat}
\end{eqnarray}
From now on, $\vec S_i$ and $\vec T_i$ are denoted by $\vec u_i$ in the uniform fashion.
A thermal distribution of $\vec u_i$ is
described by the distribution function: 
\begin{eqnarray}
w_i^u (\vec u_i) = {1 \over \nu_i^u} 
\exp \Bigl (\vec \lambda_i^u \cdot {\vec u_i \over |\vec u_i|} \Bigr ),
\label{eq:distribution}
\end{eqnarray}
where $\vec \lambda_i^u$ is the mean field and 
$\nu_i^u$ is the normalization factor given by 
\begin{eqnarray}
\nu_i^s = \int_0^\pi d \theta_i^s \int_0^{2 \pi} d \phi_i^s \sin \theta_i^s
\exp (|\vec \lambda_i^s| \cos \theta_i^s),
\label{eq:norms}
\end{eqnarray}
and
\begin{eqnarray}
\nu_i^t = \int_0^{2 \pi} d \theta_i^t
\exp (|\vec \lambda_i^t| \cos \theta_i^t).
\label{eq:normt}
\end{eqnarray}
The free energy is given by a summation 
of the expectation value of the Hamiltonian 
and entropy:    
\begin{eqnarray}
{\cal F} =\langle {\cal H} \rangle_{st} - T N({\cal S}^s + {\cal S}^t),
\label{eq:free}
\end{eqnarray}
where $N$ is the number of Mn ions
and ${\cal S}^u$ is the entropy for $\vec u$ defined by
\begin{eqnarray}
{\cal S}^u = -\langle \ln w^u (\vec u) \rangle_u. 
\label{eq:entropy}
\end{eqnarray}
$\langle A \rangle _{u}$ implies the expectation value of $A$ 
with respect to the distribution function. 
$M_u = \langle \vec u \cdot \vec \lambda^u /(|\vec u| |\vec \lambda^u|) \rangle_u$ 
is adopted as an order parameter in 
$\langle {\cal H}_J \rangle_{st}$, 
$\langle {\cal H}_H \rangle_{st}$ and 
$\langle {\cal H}_{AF} \rangle _{st}$. 
The relation $3 \langle \vec S_i \rangle_s = \langle \vec S _{t_{2g} i} \rangle_s$ 
is assumed. 
As for $ \langle {\cal H}_t \rangle_{st}$, 
the rotating frames in the spin and pseudo-spin spaces are introduced. 
The electron annihilation operator is decomposed as 
$\widetilde d_{i \gamma \sigma}= z_{i \sigma}^s z_{i \gamma}^t h_i^\dagger$  
where $h_i^\dagger$ is the creation operator of a spin- and orbital-less fermion 
describing a hole motion and
$z_{i \sigma(\gamma)}^{s(t)}$ is an element of the unitary matrix 
in the spin (pseudo-spin) frame. \cite{ishihara2,ishihara3}  
These are given by 
$z_{i \uparrow}^s = \cos (\theta^s_i / 2) e^{-i\phi^s_i / 2}$, 
$z_{i \downarrow}^s = \sin (\theta^s_i / 2) e^{i\phi^s_i / 2}$, 
$z_{i a}^t = \cos (\theta^t_i / 2)$ and 
$z_{i b}^t = \sin (\theta^t_i / 2)$.
${\cal H}_t$ is rewritten as
\begin{eqnarray}
{\cal H}_t = \sum_{\langle i j \rangle \sigma \gamma \gamma'}
(z_{i \sigma}^{s *} z_{j \sigma}^s )
(z_{i \gamma}^{t *} t_{ij}^{\gamma \gamma'} z_{j \gamma'}^t) 
h_i h_j^\dagger 
+ H.c. , 
\label{eq:htzz}
\end{eqnarray}
and is diagonalized in the momentum space 
as follows 
\begin{eqnarray}
\langle {\cal H}_t \rangle_{st} = 
\biggl \langle
\sum_{\vec k} \sum_{l=1}^{N_l}
\varepsilon_{\vec k}^l f_F ( \varepsilon_{\vec k}^l - \varepsilon_F )
\biggr \rangle_{st}\, ,  
\label{eq:energyt}
\end{eqnarray}
where 
$\varepsilon_{\vec k}^l$  is the energy in the $l$-th band for   
the spin- and orbital-less fermion with momentum $\vec k$ and $N_l$ is the number of the band. 
The Fermi energy $\varepsilon_F$ is determined by 
the condition 
$x = (1/N) \sum_{\vec k} \sum_{l} f_F ( \varepsilon_{\vec k}^l - \varepsilon_F )$ 
where 
$f_F(\varepsilon)$ is the Fermi distribution function. 
The mean field solutions are obtained by minimizing $\cal F$ with respect to 
$\vec \lambda^{u}_i$. 
\section{Numerical results}
\subsection{spin and orbital phase diagram at finite temperature }
\label{sub:diagram}
%
%
% fig1
%
\begin{figure}
\epsfxsize=0.7\columnwidth
\centerline{\epsffile{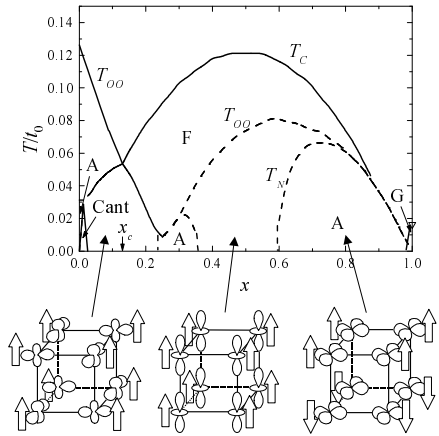}}
%
%\centerline{\BoxedEPSF{fig1.eps scaled 450}}
%\vspace{2mm}
%
\caption{The spin and orbital phase diagram as a function  of 
hole concentration $(x)$ and temperature $(T)$. 
The schematic orbital states are also shown.
$T_C$, $T_N$ and $T_{OO}$ indicate ferromagnetic Curie temparature, N$\rm{\acute e}$el temperature 
and orbital ordering temperature, respectively. 
$F$, $A$ and $G$ indicate ferromagnetic, A-type antiferromagnetic and 
G-type antiferromagnetic phases, respeceively. 
The solid (broken) lines is for the second (first) order phase transition. 
The parameter values are chosen to be 
$J_1/t_0=0.25$, $J_2/t_0=0.075$, and $J_{AF}/t_0=0.0035$.}
\label{fig:fig1}
\end{figure}
%
% fig1
%
%
The spin and orbital states 
at finite $T$ and $x$ are numerically calculated by utilizing the mean field theory 
introduced in the previous section.  
The four types of spin structure, that is, 
the ferromagnetic (F) structure and layer (A)-type, rod (C)-type 
and NaCl (G)-type AF ones are considered. 
As for the orbital,  
the ferromagnetic-like structure where one kind of orbital exists  
and the G-type one where two kinds of orbital 
are alternately aligned in the [111] direction are considered. 
The numerical results are shown in Fig.~\ref{fig:fig1}. 
Parameter values are chosen to be $J_1/t_0 = 0.25, J_2/t_0 = 0.075$ and $J_{AF}/t_0 = 0.0035$.  
The schematic pictures of the orbital ordered states are also presented. 
With increasing $x$ from $x=0$, 
the spin structure changes as A-type AF 
$\rightarrow$ F $\rightarrow$ A-type AF
$\rightarrow$ G-type AF 
which is associated with 
change of the orbital states;   
in the region of $x \le 0.25$, 
the interaction caused by ${\cal H}_J$
is the dominant one between nearest neigboring orbitals and 
the G-type AF orbital ordered state is brought about. 
The type of the orbital is denoted as 
$(\Theta_A^t/ \Theta_B^t) = ({\pi \over 2} / -{\pi \over 2})$ 
where $\Theta_{A(B)}^t$ is the angle of the pseudo-spin in sublattice $A (B)$ 
and its definition is the same with $\theta_i^t$ in Eq.~(\ref{eq:thetat}). 
These orbitals are mixtures of 
$d_{3x^2-r^2}$ and $d_{y^2-z^2}$, and $d_{3y^2-r^2}$ and $d_{z^2-x^2}$, 
respectively.  
Above $x = 0.25$, the F-type orbital ordered state is realized. 
In particular, 
the $d_{x^2-y^2}$ orbital is uniformly aligned 
in the A-type AF spin phase above $x=0.6$. 
A large hopping integral for electrons in the $xy$ plane 
in this orbital ordered state 
is energetically favored in the A-type AF state 
where the hopping in the $z$ direction is prohibited. \cite{maezono}
The calculated results of spin and orbital phase 
diagram at $T = 0$ are consistent with those obtained 
by the Hartree-Fock theory. \cite{maezono}
The spin and orbital phase diagram at finite $T$ was 
calculated in Ref.~\onlinecite{sheng} 
where the Monte Carlo method in a finite-size cluster was used 
in the spin-orbital-lattice coupled model. 
\par
Now, let us focus on the spin ordering temperatures, i.e. $T_C$ and $T_N$, 
and the orbital ordering one, $T_{OO}$, in Fig.~\ref{fig:fig1}.  
These ordering temperature vs. $x$ curves qualitatively reproduce   
the experimental results observed in $\rm La_{1-\it x}Sr_{\it x}MnO_3$,
\cite{tokura2,akimoto} $\rm Pr_{1-\it x}Sr_{\it x}MnO_3$\cite{tokura2,tomioka2}  
and Nd$_{1-x}$Sr$_x$MnO$_3$ \cite{kuwahara} except for the narrow region of the charge ordered phase  
in Nd$_{0.5}$Sr$_{0.5}$MnO$_3$. 
It is shown in Fig.~1 that 
$T_{OO}$ is higher (lower) than $T_C$ 
in the region of $x<0.1$ ($x>0.1$). 
The dominant interaction in the region of $x<0.1$ is 
provided by ${\cal H}_J$ where  
the effective interaction between orbitals in the paramagnetic state 
and that between spins in the orbital disordered state are
given by $3J_1/2$ and $J_1/2$, respectively.  
Here, the first term in ${\cal H}_J$ is considered. 
Thus, $T_{OO}$ is higher than $T_C$. 
On the other hand, 
in the region where ${\cal H}_t$ is dominant, 
gain of the kinetic energy associated with the long range ordering 
causes the transition; 
it is assumed that doped holes are introduced at the bottom of the band 
and the kinetic energy is proportional to the band width. 
The ratio of the band width in the ferromagnetic state 
to that in the paramagnetic one is obtained as
$3/2$ 
where the orbital disordered state is assumed. 
On the other hand, 
the ratio of the band width in the F-type orbital ordered state 
to that in the orbital disordered state
is obtained as $\pi^2/8$ 
where the spin paramagnetic state is assumed.
The energy gain associated with the orbital ordering is 
smaller than that with the spin one,  
so that $T_C$ is higher than $T_{OO}$.
This is attributed to the hopping integral 
between different kinds of orbital. 
\par
Between $T_N$ for the A-type AF state and $T_{OO}$, 
the relation $T_N \le T_{OO}$ is satisfied in the whole region of $x$ 
in Fig.~\ref{fig:fig1}. 
In addition, the orbital order-disorder transition 
and the A-type AF one are of the first order   
in the region of $x > 0.25$. 
This is numerically confirmed by discontinuity 
in the orbital (spin) order parameter at $T_{OO}$ ($T_N$). 
Both the two results originate from the anisotropy in the 
pseudo-spin space as 
discussed latter in more detail. 
\subsection{spin and orbital phase transitions in lightly doped region}
In the lightly doped region in Fig.~1,
$T_C$ $(T_{OO})$ increases (decreases)  
with increasing $x$ from $x=0$ and 
the two transition temperatures cross with each other around $x=0.13$ termed $x_c$. 
Around $x_c$, 
the coupling between spin and orbital degrees of freedom
brings about the unique phase transition as follows. 
%
%
% fig2
%
\begin{figure}
\epsfxsize=0.7\columnwidth
\centerline{\epsffile{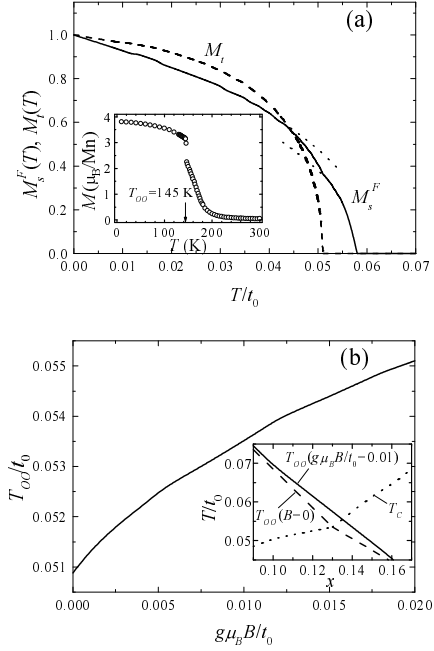}}
%
%\centerline{\BoxedEPSF{fig2.eps scaled 450}}
%\vspace{2mm}
%
\caption{(a) Temperature dependence of the magnetization $M_s^F$ (solid line) and 
the orbital order parameter $M_t$ (broken line) at $x = 0.14$. 
The inset shows the magnetization curve in $\rm La_{0.88}Sr_{0.12}MnO_3$.\cite{nojiri}
(b) Magnetic field dependence of the orbital ordering temperature $T_{OO}$ at $x=0.14$.
The inset shows $T_{OO}$ at $g \mu_B B/t_0=0$ and $0.01$ 
and the ferromagnetic Curie temperature $T_C$ around $x_c$. 
Parameter values in the calculation are the same as those in Fig.~\ref{fig:fig1}.}
\label{fig:fig2}
\end{figure}
%
% fig2
% 
%
Let us focus on the region where $x$ is slightly higher than $x_c$. 
With decreasing $T$, the system changes from the paramagnetic 
phase with the orbital disordered state to 
the ferromagnetic phase and then to the ferromagnetic one with 
the orbital ordered state. 
The temperature dependence of the magnetization at $x=0.14$ is presented in Fig.~2 (a) 
where the order parameter of the orbital ordered state is also plotted. 
It is shown that the magnetization is enhanced below $T_{OO}$. 
This originates from the coupling between spin and orbital in ${\cal H}_{J}$ (Eq.~(\ref{eq:hj})); 
in the mean fields theory, 
the effective interaction between nearest neighboring spins is given by 
$-2J_1(\langle n_i n_j \rangle/4-\langle \tau^l_i \tau^l_j\rangle)$ where 
the first term in ${\cal H}_J$ is considered. 
With taking into account the fact that the orbital ordered state is $(\Theta_A,\Theta_B)=(\pi/,3\pi/2)$ 
in this region of $x$, 
the effective interaction is rewritten as 
$-2J_1(1-x)^2(1/4+3M_t^2/16)$ for $l=x$ and $y$ and 
$-2J_1(1-x)^2/4$ for $l=z$. 
The effective magnetic interaction 
and thus the the magnetization increase below $T_{OO}$. 
In the opposite way, 
the orbital order-disorder transition 
is influenced by change of the spin state. 
The magnetic field dependence of $T_{OO}$ at $x=0.14$  
is presented in Fig.~2 (b).  
The inset shows change of the phase diagram by applying the magnetic field around $x_c$. 
$T_{OO}$ increases with applying a magnetic field. 
The effective interaction between nearest neighboring orbitals 
is given by $2J_1(1-x)^2(3/4+M_s^2/4)$ which increases by applying the magnetic field. 
This implies that the orbital state is controlled by the magnetic field, 
although the field is not a canonical external field for the pseudo-spin. 
\par
This unique phase transition originating from the coupling 
between spin and orbital is observed in manganites. \cite{endoh,nojiri}
In La$_{0.88}$Sr$_{0.12}$MnO$_3$, 
the ferromagnetic ordering occurs at 175K and  
the orbital ordering is confirmed below 145K by the resonant x-ray scattering 
which is a direct probe to detect the orbital ordering. 
The ferromagnetic phase with the 
orbital disordered state changes into the phase with the 
orbital ordered state at 145K, so that it 
corresponds to the calculated phase transition at $T_{OO}$ around $x=0.14$.
It is experimentally confirmed that   
the magnetization is enhanced below 145K (the inset of Fig.~2(a)) 
and the orbital ordering temperature increases 
with applying the magnetic field. \cite{senis,ghosh,uhlenbruck,nojiri}
These experimental results are well explained by the present calculation 
and are strong evidences of the novel coupling 
between spin and orbital in this compound. 
\subsection{phase transition in highly doped region and A-type AF metal}
In this subsection, we focus on the phase transition 
in the highly doped region ($x >0.5$) in Fig.~1 where the A-type 
AF state associated with the $d_{x^2-y^2}$ 
orbital ordered one appears. 
There are two kinds of carrier concentration regions termed 
region I ($0.8>x>0.6$) where a sequential phase transition 
from the paramagnetic state to the ferromagnetic one and to the A-type AF one 
occurs with decreasing $T$, and region II ($x>0.8$) where the transition 
from the paramagnetic state to the A-type AF one occurs.
The temperature dependences of the spin order parameters at 
$x=0.725$ (region I) and $x=0.9$ (region II) are shown in Figs.~3(a) and 
(b), respectively. 
$M_s^{F}$ and $M_s^{AF}$ are the order parameters of the ferromagnetic 
and A-type AF spin structures, respectively. 
As shown in Fig.~3(a), 
$M_s^F$ appears at $T_C/t_0=0.95$ where the transition is of the second order. 
With decreasing $T$, 
the F-type orbital ordered state with $d_{x^2-y^2}$ orbital appears 
at $T_{OO}/t_0=0.72$ and the transition from the ferromagnetic phase 
to the A-type AF one occurs at $T_N$. 
This transition is of the first order and 
the canted AF phase does not appear between the ferromagnetic and A-type AF phases.  
This sequential phase transition is caused by the thermal fluctuation of the orbital;
as previously mentioned, 
the A-type AF state and the F-type orbital ordered one with 
$d_{x^2-y2}$ orbital are cooperatively stabilized at $T=0$. 
With increasing $T$, the thermal fluctuation of orbital grows up and 
the hopping integral in the $z$ direction becomes finite. 
As a result, the double exchange interaction in this direction overcomes 
the antiferromagnetic SE one and the ferromagnetic phase is stabilized. 
%
%
% fig3
%
\begin{figure}
\epsfxsize=0.7\columnwidth
\centerline{\epsffile{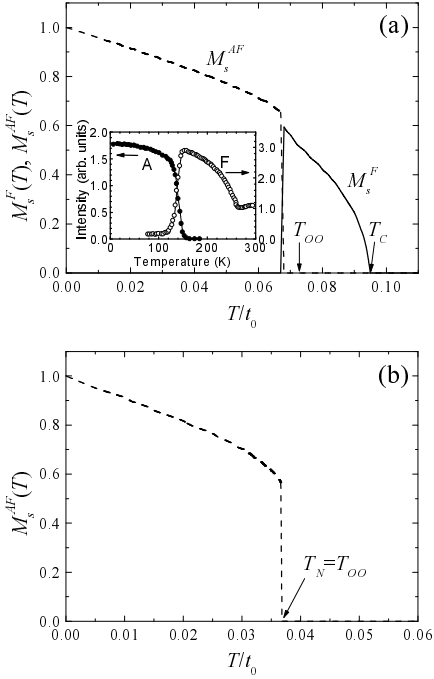}}
%
%\centerline{\BoxedEPSF{fig3.eps  scaled 550}}
%\vspace{5mm}
%
\caption{Temperature dependences of the spin order parameters  
(a) at $x = 0.725$ and (b) at $x = 0.9$.  
The solid and broken lines show the order parameters 
in the ferromagnetic state ($M_s^F$) and the A-type antiferromagnetic 
one ($M_s^{AF})$, respectively.  
Parameter values are the same as those in Fig. \ref{fig:fig1}.  
The inset in (a) shows the temperature dependences of magnetic Bragg reflections 
in $\rm Pr_{0.5}Sr_{0.5}MnO_3$.\cite{kawano}}
\label{fig:fig3}
\end{figure}
%
% fig3
%
%
In the region II, the A-type AF ordering and the F-type 
orbital one with $d_{x^2-y^2}$ orbital occur at the same 
temperature where the transition is of the first order 
as shown in Fig.~3(b).  
In both the regions I and II, the relation $T_{N} \le T_{OO}$ 
is satisfied. 
The first order transition at $T_{OO}$ and 
this relation between $T_{OO}$ and $T_N$ originate from 
the breaking of the inversion symmetry of the system with respect to the 
orbital pseudo-spin operator, as discussed in the next section. 
The calculated results of the sequential phase transition in the region I 
well reproduce the experimental results observed in  
$\rm Pr_{0.5}Sr_{0.5}MnO_3$\cite{kawano} (the inset of Fig.~\ref{fig:fig3} (a)).  
The first order transition at $T_{OO}(=T_N)$ in the region II is 
consistent with the experiments in $\rm Nd_{0.45}Sr_{0.55}MnO_3$ \cite{kawano,kuwahara} 
where the Mn-O band length in the $z$ direction 
($xy$ plane) is confirmed to become short (long) at $T_N$.\cite{kajimoto}
It implies the F-type orbital ordering with $d_{x^2-y^2}$ orbital as 
predicted from the present calculation. 
\par
\begin{figure}
\epsfxsize=0.7\columnwidth
\centerline{\epsffile{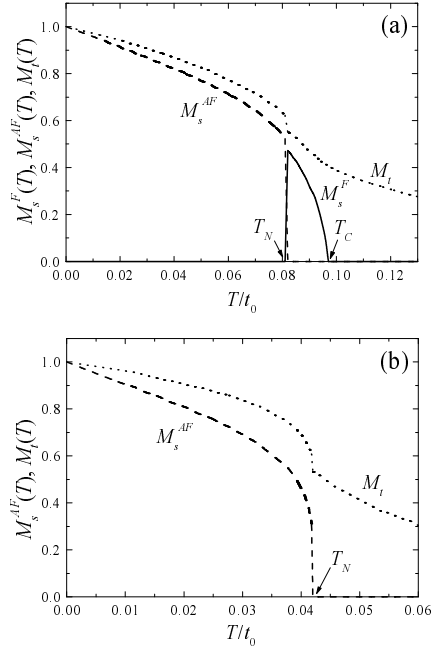}}
%
%\centerline{\BoxedEPSF{fig4.eps  scaled 550}}
%\vspace{5mm}
%
\caption{Temperature dependences of the spin and orbital order parameters 
(a) at $x = 0.725$ and (b) at $x = 0.9$ 
in the tetragonal lattice.
The solid, broken and dotted lines show the order parameters 
for the ferromagnetic structure ($M_s^F$), 
the A-type AF one ($M_s^{AF})$ and the F-type orbital ordered one ($M_t$), respectively.  
The uniaxial anisotropy is introduced as  
$t_0^{xy}/t_0^z=\sqrt{J_{AF}^{xy}/J_{AF}^{z}}=R$ with $R=1.2$
where $t_0^{xy(z)}$ and $J_{AF}^{xy(z)}$ are the hopping integral and the 
antiferromagnetic SE interaction in the $xy$ plane (the $z$ direction). 
The other parameter values are the same as those in Fig. \ref{fig:fig1}.}
\label{fig:fig4}
\end{figure}

In the actual compounds, where the A-type AF state is observed, 
the tetragonal lattice distortion is observed and 
the cubic symmetry is broken far above $T_N$. \cite{kawano,kajimoto}
We simulate this distortion by 
introducing the uniaxial anistoropy of the hopping integral and the 
SE interaction and investigate the phase transition.
It is assumed that $t_0^{xy}/t_0^z = \sqrt{J_{AF}^{xy}/J_{AF}^{z}}=R$ 
where $t_0^{xy(z)}$ and $J_{AF}^{xy(z)}$ are the hopping integral and 
the antiferromagnetic SE interaction in the $xy$ plane ($z$ direction). 
The temperature dependences of the spin and orbital order parameters at 
$x=0.725$ and $x=0.9$ with $R=1.2$ 
are shown in Figs.~\ref{fig:fig4} (a) and (b), respectively. 
Due to the uniaxial anisotropy, 
$M_t$ is finite above the orbital ordering temperature in the system 
with the cubic symmetry.
It is worth to note that 
1) $T_N$ for the A-type AF state increases and  
2) the transition at $T_N$ is of the first order, although the discontinuity of $M_s^{AF}$ is reduced.  
The latter is attributed to diminution of the change of $M_t$ at $T_N$. 
%
%
% fig4
%
%
% fig4
%
%
\par
Finally, the magnetic field dependence of $T_{OO}$ is shown in Fig.~5.  
$T_{OO}$ decreases with applying the magnetic field in the 
region of $g \mu_B B/t_0 < 0.0025$ where the 
nearest neighboring spins in the $z$ direction are canted. 
The spins become parallel at $g \mu_B B/t_0 =0.0025$ termed $B_c$, 
and $T_{OO}$ increases with increasing the magnetic field above $B_c$. 
The orbital ordered state below $T_{OO}$ is of the F-type with $d_{x^2-y^2}$ orbital 
and does not depend on magnitude of the magnetic field. 
Above and below $B_c$, the different mechanisms dominate 
the magnetic field dependence of $T_{OO}$; 
in the region of $B>B_c$, the spin canting due to the magnetic field 
promotes the electron hopping 
in the $z$ axis which weakens the orbital ordered state. 
On the other hand, above $B_c$, magnitude of   
the magnetic moment is enhanced by increasing of the magnetic field and the orbital ordered state 
associated with the ferromagnetic spin structure is stabilized. 
%
%
% fig5
%
\begin{figure}
\epsfxsize=0.7\columnwidth
\centerline{\epsffile{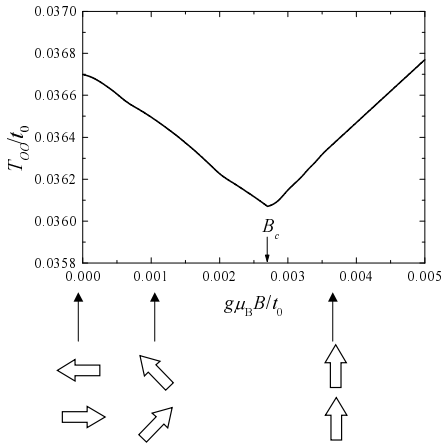}}
%
%\centerline{\BoxedEPSF{fig5.eps  scaled 550}}
%\vspace{5mm}
%
\caption{Magnetic field dependence of the orbital ordering temperature at $x=0.9$ and 
schematic spin configurations. 
Parameter values are the same as those in Fig. \ref{fig:fig1}.} 
\label{fig:fig5}
\end{figure}
%
% fig5
%
%
The magnetic field dependence of $T_N$ for the A-type AF spin structure  
is recently measured in $\rm Nd_{0.45}Sr_{0.55}MnO_3$. \cite{kuwahara}
It is experimentally shown that $T_N$ gradually decreases with increasing the magnetic field. 
From the present calculation, it is predicted that this reduction of 
$T_N$ is accompanied with that of $T_{OO}$, and 
with increasing the magnetic field furthermore, 
$T_{OO}$ increases above the critical value of the field. 
\par
\section{First order transition and orbital degree of freedom}
\label{sec:landau}
As mentioned in Sect.~III, 
several characteristics of the phase transitions in manganites 
are attributed to the unique properties of the orbital degree of freedom. 
In this section, we study analytically the phase transition
by expanding the free energy with respect to the spin and orbital 
order parameters. \cite{degennes}
Let us consider the ferromagnetic and A-type AF spin structures and the 
F and G-type AF orbital ones. 
The expectation values of the Hamiltonian at finite temperature 
are calculated by the mean field theory introduced in Sect.~II 
and are expanded with respect to $M_s$ and $M_t$.
It is assumed that doped holes are introduced at the bottom of the 
band denoted by $\varepsilon^l_{\vec k}$ in Eq.~(\ref{eq:energyt}). 
The explicit form of $\langle {\cal H}_J \rangle_{st}$ is 
given by 
\begin{eqnarray}
{\langle {\cal H}_{J} \rangle_{st} \over N} 
&=&-{(1-x)^2 \over 2}\sum_{l = x,y,z} \nonumber \\
&\times& \Bigl \{  J_1  
\! \bigl ( 3 +  \cos\Theta_l^s M_s^2 \bigr ) 
   \bigl ( 1 - A_{2l} M_t^2 \bigr ) \nonumber \\
&+& \ \ J_2 
 \bigl ( 1 - \cos\Theta_l^s M_s^2 \bigr ) \nonumber \\
&&\qquad\qquad
\times 
\bigl ( 3   + A_{1l} M_t + A_{2l} M_t^2 \bigr ) \Bigr \},
\label{eq:hjexpand}
\end{eqnarray}
where 
\begin{equation}
A_{1l}=2( C_{A l} + C_{B l} ) , 
\end{equation}
and 
\begin{equation}
A_{2l}=C_{A l} C_{B l} , 
\end{equation}
with 
$C_{A(B)l}=\cos ( \Theta_{A(B)}^t+2 \pi m_l/3 )$ and 
$(m_x,m_y,m_z)=(1,-1,0)$. 
$\Theta_l^{s}$ $(l=x,y,z)$ is the relative angle between nearest neighboring 
spins in direction $l$ and 
$\Theta_{A(B)}^{t}$ is the angle of the orbital pseudo-spin in sublattice $A(B)$. 
$\langle {\cal H}_t \rangle_{st}$ 
is proportional to the band width $W$ of the spin- and orbital-less fermion as 
\begin{equation}
{\langle {\cal H}_t \rangle_{st} \over N}
= -x{ W \over 2} , 
\label{eq:bandwdth}
\end{equation}
and is expanded with respect to $M_s$ and $M_t$ up to 
the orders of $M_s^2$ and $M_t^3$ as follows 
\begin{eqnarray}
{\langle {\cal H}_t \rangle_{st} \over N}
&=& - t_0 x {16 \over 3 \pi^2} \sum_{l = x,y,z} 
\Bigl( 1+{3 \over 5} \cos \Theta_l^s M_s^2 \Bigr) \nonumber \\
&&\;\;\;\times
\Bigl(1+\alpha_{1l} M_t + \alpha_{2l} M_t^2 +\alpha_{3l} M_t^3 \Bigr),
\label{eq:htexpand}
\end{eqnarray}
where 
\begin{equation}
\alpha_{1l}={2 \over 3} ( C_{Al}+C_{Bl}) , 
\end{equation}
\begin{equation}
\alpha_{2l}={2 \over 15} ( 1 - C_{Al}^2 - C_{Bl}^2 ) 
+ {4 \over 9} C_{Al} C_{Bl} , 
\end{equation}
and 
\begin{eqnarray}
\alpha_{3l}&=&{ 4 \over 105} ( C_{Al}^3 + C_{Bl}^3 )
-{4 \over 45} ( C_{Al}^2 C_{Bl} + C_{Al} C_{Bl}^2 ) \nonumber \\
&+&{1 \over 63} ( C_{Al} + C_{Bl}) . 
\end{eqnarray}
The detailed derivation of Eq.~(\ref{eq:htexpand}) is 
presented in Appendix.
It is worth to note that the terms 
which are proportional to $M_t$ or $M_t^3$ appear
in $\langle {\cal H}_J \rangle _{st}$ and $\langle {\cal H}_t \rangle_{st}$. 
This is because the inversion symmetry in the system 
with respect to the orbital pseudo-spin is broken, 
i.e. the free energy with $T_z=1/2$ is different from 
that with $T_z=-1/2$. 
This is highly in contrast to the spin case where 
the inversion symmetry with respect to the spin operator 
is preserved due to the time reversal symmetry in the system without magnetic field. 
Being based on Eqs.~(\ref{eq:hjexpand}) and (\ref{eq:htexpand}), 
we investigate the phase transition in the F and A-type AF spin 
structures in more detail. 
\par
{\it \underline {Ferromagnetic structure}}:  
Eqs.~(\ref{eq:hjexpand}) and (\ref{eq:htexpand}) 
with the relation $\Theta_l^s=0$ for $l=x,y$ and $z$  are given by 
\begin{eqnarray}
{\langle {\cal H}_{J} \rangle_{st} \over N} 
=&-&(1-x)^2{3\over8} J_1 \bigl( 3 + M_s^2 \bigr) \bigl( 1 + B_2 M_t^2 \bigr)\nonumber \\
 &-&(1-x)^2{3\over8} J_2 \bigl( 1 - M_s^2 \bigr) \bigl( 3 - B_2 M_t^2 \bigr), 
\end{eqnarray}
and 
\begin{eqnarray}
{\langle {\cal H}_t \rangle_{st} \over N}
= &-& t_0 x {16 \over {\pi^2}}
\Bigl( 1+{3 \over 5} M_s^2 \Bigr) \nonumber \\
&&\qquad\qquad\times
\Bigl(1 + \beta_{2} M_t^2 +\beta_{3} M_t^3 \Bigr),
\label{eq:htferro}
\end{eqnarray}
respectively, 
with 
\begin{equation}
B_2=-{1 \over 2} \cos(\Theta_A^t-\Theta_B^t) , 
\end{equation}
\begin{equation}
\beta_2 ={2 \over 9} \cos( \Theta_A^t - \Theta_B^t ) , 
\end{equation}
and 
\begin{eqnarray}
\beta_3 &=& {4 \over 105} 
 (\cos^3 \Theta_A^t +\cos^3 \Theta_B^t)   \nonumber \\  
&-&{1 \over 35} (\cos \Theta_A^t+\cos \Theta_B^t) \nonumber \\
&-&{1 \over 45} 
  \Bigl( \cos \bigl( 2\Theta_A^t + \Theta_B^t \bigr) 
        +\cos \bigl( \Theta_A^t + 2\Theta_B^t \bigr) \Bigr).
\label{eq:beta3}
\end{eqnarray}
The terms being proportional to $M_t$ vanish due to the cubic symmetry 
of the ferromagnetic spin structure. 
The coefficients ${\rm B}_2$ and $\beta_2$ become 
the largest at $\Theta_A^t=\Theta_B^t-\pi$ and 
at $\Theta_A^t=\Theta_B^t$, respectively. 
That is, ${\cal H}_J$ and ${\cal H}_t$ favor AF- and F-type orbital ordered states, 
respectively. 
\par
Let us focus on the term being proportional to $M_t^3$ 
in $\langle {\cal H}_t \rangle_{st}$. 
In the case where this term is relevant, 
the orbital order-disorder transition is of the first order 
according to the Landau criterion in the phase transition. 
It corresponds to the transition at $T_{OO}$ 
in the region of $x>0.25$ in Fig.~1 and 
is consistent with the first order transitions at the orbital ordering temperature  
observed in several manganites. 
On the other hand, 
the transition in the region of $x<0.25$ in Fig.~1 
is of the second order. 
This is because 1) the term being proportional to $M_t^3$ 
does not appear in $\langle {\cal H}_J \rangle_{st} $, and 
2) in this hole concentration region, the G-type orbital ordered state with 
$\Theta_A^t=\Theta_B^t-\pi$ is realized. 
The inversion symmetry with respect to the pseudo-spin operator is preserved, 
i.e. $\beta_3=0$ in Eq.~(\ref{eq:beta3}) in this orbital ordered state.  
The phase transition in this hole concentration region is discuss in more detail in Sect.~V. 
The first order transition in the cooperative Jahn-Teller system 
and its relation to the terms proportional to $Q^3$, where $Q$ indicates 
the normal mode of a MnO$_6$ octahedron, was discussed in Ref. \onlinecite{kataoka}. 
\par
The parameter $\beta_3$ becomes the largest at 
$\Theta_A^t=\Theta_B^t=(2n+1)\pi/3$ with $n=(1,2,3)$ and 
determines the orbital state uniquely.  
With taking into account $\beta_2$ together with $\beta_3$, 
$\langle {\cal H}_t \rangle_{st}$ favors 
the F-type orbital ordered state with 
$d_{x^2-y^2}$, $d_{y^2-z^2}$ and $d_{z^2-x^2}$ (the so-called leaf-type orbital) 
rather than 
$d_{3z^2-r^2}$, $d_{3x^2-r^2}$ and $d_{3y^2-r^2}$ (the so-called ciger-type one). 
Since the band width in the F-type orbital ordered state at $T=0$ 
does not depend on types of the orbital, 
the thermal fluctuation stabilises the leaf-type orbital;  
the F-type orbital ordered state with $d_{x^2-y^2}$ $(d_{3z^2-r^2})$ 
mixes with the AF-type one with $(d_{3x^2-r^2}/d_{3y^2-r^2})$
($(d_{y^2-z^2}/d_{z^2-x^2})$) through the thermal fluctuation.  
The band width in the $(d_{3x^2-r^2}/d_{3y^2-r^2})$ state 
is $5t_0$ which is larger than that in the $(d_{y^2-z^2}/d_{z^2-x^2})$ 
one ($3t_0$). 
The smaller band width in the latter state is attributed 
to the fact that the hopping integral in the $xy$ plane is zero 
in this orbital state. 
\par
{\it \underline{A-type AF structure}}: 
In this spin structure,  
$\langle {\cal H}_J \rangle_{st}$ 
and 
$\langle {\cal H}_t \rangle_{st}$ 
with $\Theta_{x}^s=\Theta_y^s$=0 and 
$\Theta_z^s=\pi$ are given by  
\begin{equation}
{\langle {\cal H}_{J} \rangle_{st} \over N}=(1-x)^2{1 \over 16}(J_1 D_1+J_2 D_2) , 
\end{equation}
with 
\begin{eqnarray}
D_1&=&-18-2M_s^2 
+9 \cos(\Theta_A^t-\Theta_B^t) M_t^2 \nonumber \\ 
&+&\Bigl ( 3 \cos(\Theta_A^t-\Theta_B^t)-2\cos \Theta_A^t \cos \Theta_B^t
\Bigr ) M_s^2M_t^2 , 
\end{eqnarray}
and 
\begin{eqnarray}
D_2&=&-18+6 M_s^2
-3 \cos(\Theta_A^t-\Theta_B^t) M_t^2 \nonumber \\
&-&4\bigl( \cos\Theta_A + \cos\Theta_B \bigr) M_s^2M_t  \nonumber \\
&+&\Bigl ( 3 \cos(\Theta_A^t-\Theta_B^t)-2\cos \Theta_A^t \cos \Theta_B^t
\Bigr ) M_s^2M_t^2 ,
\end{eqnarray}
and  
\begin{eqnarray}
{\langle {\cal H}_t \rangle_{st} \over N}
&=&- t_0 x 
\biggl({16 \over {\pi^2}}\biggr) \nonumber \\
&\times& \Bigl( 1 + {1 \over 5}  M_s^2 
 + \gamma_1 M_s^2 M_t + \beta_{2} M_t^2+\beta_3 M_t^3 \Bigr), 
\label{eq:htatype}
\end{eqnarray}
with 
\begin{eqnarray}
\gamma_1 &=& - {4 \over 15} \bigl( \cos \Theta_A^t + \cos \Theta_B^t \bigr) , 
\end{eqnarray}
respectively. 
The most remarkable difference of the results in the A-type AF structure  
from that in the ferromagnetic one is the terms being proportional 
to $M_s^2 M_t$. 
The origin of these terms is the 
anisotropic spin structure in the A-type AF state 
which breaks the cubic symmetry in the system. 
Because of these terms, the order parameter of the A-type AF state acts as a 
magnetic field on the orbital pseudo-spin space and 
the relation $T_{OO} \ge T_N$ is derived. 
This relation is seen in the present phase diagram (Fig.~1)  
and also in the experimental results in several manganites. 
\par
Now we focus on these terms being proportional to $M_tM_s^2$. 
At $x=0$ where ${\cal H}_J$ is dominant, 
the AF-type orbital ordered states with $(\Theta^t /\Theta^t+\pi)$ for any $\Theta^t$
is realized above $T_N$. 
Below $T_N$, the term being proportional to $M_s^2 M_t$ becomes relevant 
and the orbital state is uniquely determined as 
$(\Theta^t/-\Theta^t)$ with 
\begin{eqnarray}
\Theta^t=\cos^{-1}
{4J_2M_s^2 
\over 
3(3J_1-J_2)M_t+(J_1+J_2)M_s^2M_t} . 
\label{eq:orbcant}
\end{eqnarray}
With decreasing temperature below $T_N$, 
$\Theta^t$ continuously changes from $\pi/2$ at $T_N$ to
$\cos^{-1}\{ 2J_2/(5J_1-J_2) \}$ at $T=0$.
This orbital state favors the 
anitiferromagnetic interaction in the $z$ direction 
in this spin structure. 
In the highly doped region where ${\cal H}_t$ is dominant,   
the term $\gamma_1 M_s^2 M_t$ in Eq.~(\ref{eq:htatype}) 
favors the F-type orbital ordered state with $\Theta_A^t=\Theta_B^t=\pi$ 
which corresponds to the state with $d_{x^2-y^2}$, as mentioned 
in the previous section. 
The terms being proportional to $M_s^2 M_t$ 
also appear in the C-type AF spin structure where 
the relation $T_{OO}  \ge T_N$ is derived. 
The coefficients of the term in 
$\langle {\cal H}_J \rangle_{st}$ and $\langle {\cal H}_t \rangle_{st}$ 
are given by $J_2(1-x)^2(\cos \Theta_A^t+\cos \Theta_B^t)/2$
and 
$-t_0x32  (\cos \Theta_A^t+\cos \Theta_B^t)  /(5\pi^2)$ 
which favor  
the G-type with $(d_{3z^2-r^2}/ d_{x^2-y^2})$ 
and F-type with $d_{3z^2-r^2}$ orbital ordered states, respectively. 
\section{Summary and discussion}
In this paper, we study roles of the orbital degree of freedom in 
phase transition in perovskite mangnanites.  
The effective Hamiltonian which includes the orbital degree of freedom 
as well as the spin and charge ones is utilized and the mean field 
theory at finite temperature and carrier concentration is adopted. 
Through both the numerical and analytical calculations  
based on this theory, it is revealed that 
several characteristics of the phase transition 
observed in manganites originate from 
the unique properties of the orbital degree of freedom.
The obtained results are summarized as follows: 
1) The orbital order-disorder transition is of the first order in the wide region of $x$, 
and $T_N$ for the anisotropic spin structure, 
such as the A- and C-type AF ones, is lower than $T_{OO}$. 
Both the results originate from the fact that the inversion symmetry in the system  
is broken with respect to the orbital pseudo-spin operator 
and the terms being proportional to $M_s^2 M_t$ and $M_t^3$ exist in the free energy. 
These results are consistent with the phase transition 
observed in Pr$_{0.5}$Sr$_{0.5}$MnO$_3$ and Nd$_{0.45}$Sr$_{0.55}$MnO$_3$ \cite{kawano,kuwahara,kajimoto}
where the A-type AF state with $d_{x^2-y^2}$ orbital ordered one appears.  
The calculated results may be also applicable to the orbital ordering associated with the CE-type 
AF one which is experimentally confirmed to be of the first order transition 
in Pr$_{1-x}$Ca$_{x}$MnO$_{3}$.\cite{tomioka3} 
In the present calculation, 
the phase transition at $T_{OO}$ in the region of $x<0.25$ 
is of the second order as shown in Fig.~1. 
With taking into account the higher order coupling 
between the pseudo-spin and the Jahn-Teller type distortion in a MnO$_6$ 
octahedron and the anharmonic term of the potential energy for the lattice distortion, 
the phase transition changes from the second order 
to the first one. \cite{kataoka}
However, since the phase transition experimentally observed 
in LaMnO$_3$ at 780K is almost of the second order, \cite{rodriguez,murakami2}
it is supporsed that the effects are small or are canceled out with each other 
in the compound. 
2) The relation $T_{C}>T_{OO}$ is satisfied in highly hole doped region ($x>0.1$). 
This is because gain of the kinetic energy of electrons accompanied with the 
orbital ordering is lower than that with the ferromagnetic one 
due to the hopping integral between different kinds of orbital unlike spin case.  
3) In the region where $T_{OO}$ and $T_C$ are close with each other, 
the novel phase transition is brought about due to 
the coupling between the spin and orbital degrees of freedom. 
The magnetization is enhanced below $T_{OO}$ and 
$T_{OO}$ increases by appling the magnetic field. 
These results well explain the unique experimental results 
observed in La$_{0.88}$Sr$_{0.12}$MnO$_3$. 
4) The sequential phase transition from the A-type AF phase to the 
ferromagnetic one with increasing $T$ is caused by the 
thermal fluctuation of the orbital from the $d_{x^2-y^2}$ orbital ordered states. 
The ferromagnetic interaction in the $z$ axis becomes finite due to the orbital fluctuation.  
\acknowledgments
The authors would like to thank Y. Endoh, K. Hirota, H. Nojiri and Y. Murakami 
for their valuable discussions. 
This work was supported by 
CREST (Core Research for Evolutional Science and Technology Corporation) Japan, 
NEDO (New Energy Industrial Technology Development Organization) Japan, 
and Grant-in-Aid for Scientific Research Priority Area from the Ministry of Education, 
Science and Culture of Japan.  
S. O. acknowledges the financial support of JSPS Research Fellowship for Young Scientists.  
Part of the numerical calculation was performed in 
the HITACS-3800/380 supercomputing facilities in IMR, Tohoku University. 
\appendix
\section*{expansion of $\langle {\cal H}_{\lowercase{t}} \rangle_{\lowercase{st}}$ } 
In this appendix, 
we present derivation of Eq.~(\ref{eq:htexpand}), i.e. 
the expansion of $ \langle {\cal H} _t \rangle_{st}$ 
with respect to the spin and orbital order parameters. 
We start from Eq.~(\ref{eq:bandwdth})
where 
the band width $W$ is calculated from Eq.~(\ref{eq:htzz}) as follows 
\begin{eqnarray}
W = 2 \sum_{\delta} I_{\delta}^s I_{\delta}^t , 
\label{eq:bandwidth} 
\end{eqnarray}
where 
\begin{eqnarray}
I_{ \delta}^s = \sum_{ \sigma }
\bigl \langle \bigl | z_{i \sigma}^{s *} z_{j \sigma}^s \bigr | \bigr \rangle_s , 
\end{eqnarray}
and 
\begin{eqnarray}
I_{ \delta}^t = \sum_{ \sigma }
\bigl \langle \bigl | z_{i \gamma}^{t *} t_{ij}^{\gamma \gamma'} 
                      z_{j \gamma'}^t \bigr | \bigr \rangle_t . 
\end{eqnarray}
Here, $\delta$ indicates a vector connecting site $i$ and it nearest neighboring 
site $j$. 
The spin part $I_{\delta}^s$ has the same form with Eq.~(27) in Ref. \onlinecite{degennes} 
and is given by 
\begin{eqnarray}
I_{\delta}^s={2 \over 3} \Bigl(1+ {3 \over 5}\cos \Theta_{\delta }^s M_s^2 \Bigr) \; ,
\end{eqnarray}
where $\Theta_{\delta}^s$ is the relative angle between spins 
at site $i$ and site $j$ and the relation 
$M_s=\lambda^s/3+O(\lambda^{s3})$ is used.
As for the orbital part, 
we present the derivation of $I_{\delta }^t$
with $\delta =\pm a \hat z$ termed $I_z^t$ where $a$ and $\hat z$ 
indicate a cell parameter of the cubic perovskite lattice 
and the unit vector in the $z$ direction, respectively. 
$I_{\delta}^t$ with $\delta=\pm a {\hat x} ({\hat y})$  
is given by $I_{z}^t$ where $\Theta_i^t$ is replaced by 
$\Theta_i^t+2\pi/3$ $(\Theta_i^t-2\pi/3)$. 
$I_{z}^t$ is calculated as 
\begin{eqnarray}
I_{z}^t
&=&t_0 \biggl \langle \biggl | \cos{\theta_i^t \over 2} \cos{\theta_{j}^t \over 2} 
\biggr | \biggr \rangle_t \nonumber \\
&=&{t_0 \over \nu^{t 2}}
\int_0^{2\pi} d\delta\theta_i
\int_0^{2\pi} d\delta\theta_j
e^{\lambda^t(\cos\delta\theta_i+\cos\delta \theta_j)}\nonumber \\
& \times& \biggl |\cos{\Theta_i^t + \delta \theta_i \over 2} 
\cos{\Theta_{j}^t + \delta\theta_j \over 2} \biggr |,
\label{eq:bandt1}
\end{eqnarray}
where $\delta\theta_i=\theta_i^t-\Theta_i^t$. 
The right hand side in Eq.~(\ref{eq:bandt1}) is expanded with respect to $\lambda^t$ 
up to the order of $O(\lambda^{t3})$ as follows 
\begin{eqnarray}
I_{z}^t&=& {t_0 \over (2\pi)^2} \Bigl( 1-{\lambda^{t2} \over 4} \Bigr)^2 \nonumber \\
       &\times& \biggl( \zeta_{i0}+\zeta_{i1} \lambda^t+{\zeta_{i2} \over 2!} \lambda^{t2}+
                       {\zeta_{i3} \over 3!}\lambda^{t3} \biggr)              \nonumber \\
       &\times& \biggl( \zeta_{j 0}+\zeta_{j 1}\lambda^t+{\zeta_{j 2} \over 2!}\lambda^{t2}+
                       {\zeta_{j 3} \over 3!}\lambda^{t3} \biggr) ,  
\label{eq:bandt2} 
\end{eqnarray}
where 
$\zeta_{ i n}$ ($n=0 \sim 3$ ) are given by 
\begin{equation}
\zeta_{i0}=4 , 
\end{equation}
\begin{equation}
\zeta_{i1}={4 \over 3 } \cos \Theta_i^t , 
\end{equation}
\begin{equation}
\zeta_{i2}={4 \over 15} (8-\cos^2 \Theta_i^t) , 
\end{equation}
and 
\begin{equation}
\zeta_{i3}={4 \over 35} (\cos^3 \Theta_i^t+8\cos \Theta_i^t) .  
\end{equation}
The relation 
$\nu^t = 2\pi(1+{1\over4}\lambda^{t2})+O(\lambda^{t4})$ 
is used. 
In the G-type orbital ordered state considered in Sect.~III, 
$\Theta_i^t$ in the above formulas is replaced by $\Theta_{A(B)}^t$, 
when site $i$ belongs to the orbital sublattice $A(B)$. 
By utilizing the relation $M_t=\lambda^t/2 -\lambda^{t3}/16 +O(\lambda^{t5})$, 
Eq.~(\ref{eq:htexpand}) is derived. 

\end{document}